\documentclass[%
 reprint,
superscriptaddress,
 amsmath,amssymb,
 aps, pre
]{revtex4-1}

\usepackage{graphicx}% Include figure files
\usepackage{dcolumn}% Align table columns on decimal point
\usepackage{bm}% bold math
\usepackage{hyperref}% add hypertext capabilities
\usepackage{subfig}
\usepackage{color}
\newcommand{\changes}[1]{\textcolor{black}{#1}}
\newcommand{\changesFig}[1]{{#1}}

\begin{document}

\preprint{APS/123-QED}

%\title{Capillary waves in viscous fluids}
\title{\changes{Frequency dispersion of small-amplitude} capillary waves in
viscous fluids}

\author{Fabian Denner}
\email{f.denner09@imperial.ac.uk}
\affiliation{Department of Mechanical Engineering, Imperial College London,
London, SW7 2AZ, United Kingdom}

\date{July 19, 2016}

\begin{abstract}
This work presents a detailed study of the dispersion of capillary waves with
small amplitude in viscous fluids using an analytically derived solution to the
initial value problem of a small-amplitude capillary wave as well as direct
numerical simulation. A rational parametrization for the dispersion of
capillary waves in the underdamped regime is proposed, including predictions for the wavenumber
of critical damping based on a harmonic oscillator model. The scaling resulting
from this parametrization leads to a self-similar solution of the frequency
dispersion of capillary waves that covers the entire underdamped regime, which allows an accurate evaluation of the
frequency at a given wavenumber, irrespective of the fluid properties. This
similarity also reveals characteristic features of capillary waves, for instance
that critical damping occurs when the characteristic timescales of dispersive
and dissipative mechanisms are balanced.
In addition, the presented results suggest that the widely adopted hydrodynamic
theory for damped capillary waves does not accurately predict the dispersion
when viscous damping is significant and a new definition of the damping rate,
which provides consistent accuracy in the underdamped regime, is presented.
% \begin{description} \item[Usage] Secondary publications and information
% retrieval purposes.
% \item[68.15.+e] May be entered using the \verb+\pacs{#1}+ command.
% \item[Structure] You may use the \texttt{description} environment to structure
% your abstract; use the optional argument of the \verb+\item+ command to give
% the category of each item.
% \end{description}
\end{abstract}

\pacs{Valid PACS appear here}% PACS, the Physics and Astronomy
\maketitle

\section{Introduction}
Waves at fluid interfaces are ubiquitous in two-phase flows across a wide range
of scales, from the tidal wave with a wavelength of $\lambda \sim 10^7 \,
\mathrm{m}$ and tsunamis ($\lambda > 10^5 \, \mathrm{m}$) down to wavelengths of
the order of the size of individual molecules. For interfacial waves with long
wavelength, $\lambda \gg \sqrt{\sigma/(\rho g)}$ (where $\sigma$ is the surface
tension coefficient, $\rho$ is the fluid density and $g$ is the gravitational
acceleration), so-called {\em gravity waves}, gravity is the main mechanism
governing these waves, whereas for interfacial waves with small wavelength,
$\lambda \ll \sqrt{\sigma/(\rho g)}$, so-called {\em capillary waves}, surface
tension is the dominant dispersive and restoring mechanism. In addition, viscous
stresses act preferably at small scales \citep{Lamb1932}, leading to an
increasing viscous attenuation of capillary waves with decreasing wavelength
and a decreasing frequency of capillary waves for increasing viscosity
\citep{Levich1962}.
\citet{Longuet-Higgins1992} elegantly summarized the governing mechanisms for
capillary waves: ``At small scales, the role of surface tension and viscosity
are all-important''.

Capillary waves play an important role in the capillary-driven breakup
(Rayleigh-Plateau instability) of liquid jets and ligaments \citep{Eggers2008,
Castrejon-Pita2015}, the atomization of liquid jets \citep{Tsai1997} as well as
the stability of liquid and capillary bridges \citep{Hoepffner2013} and of
liquid curtains \citep{Lhuissier2016}.
% and in the locomotion of water striders \citep{Hu2003,Hu2010}.
Capillary waves are observed at the front of short gravity waves
\citep{Longuet-Higgins1963,Longuet-Higgins1964,Longuet-Higgins1995,Crapper1970},
for instance in the ocean where they enhance the heat and mass transfer between
water and atmosphere \citep{Witting1971,Szeri1997}. Capillary waves have also
been identified as the key mechanism governing the formation of bound states of
solitary waves in falling liquid films \citep{Pradas2011} and have been observed
to enhance film thinning between two approaching interfaces, for instance in
foams and emulsions \citep{Ruckenstein1987}.
\citet{Shats2010} observed the formation of capillary rogue waves in
experiments, meaning capillary waves can serve as a prototype to study the
formation of rogue waves in a relatively easily accessible laboratory
experiment. Similarly, capillary waves are used to study wave turbulence
experimentally \citep{Falcon2007,Xia2010,Deike2012} as well as numerically
\citep{Pushkarev1996,Deike2014}. In cell biology, capillary waves influence
the behavior and properties of lipid membranes, micelles and vesicles
\citep{Crawford1987,Lebedev1989,Gompper1992}, and with respect to microfluidic
applications, capillary waves are of central interest to applications such as
surface wave acoustics \citep{Yeo2014}, microstreaming \citep{Chindam2013} and
ultrasound cavitation \citep{Tandiono2010}.

Since capillary waves inherently have a short wavelength, understanding the
physical processes and optimizing the engineering applications named in the
examples above are dependent, among others, on a detailed knowledge of the
influence of viscous stresses on capillary waves. For instance, viscosity
reduces the growth rate of the Rayleigh-Plateau instability in liquid jets
\citep{Weber1931,Chandrasekhar1961} and alters the size, number and speed of
capillary waves in between interacting solitary waves on falling liquid films
\citep{Kalliadasis2012, Dietze2016, Denner2016}.
Due to the significantly higher dissipation of capillary waves compared to
gravity waves, capillary waves occurring on the forward facing slope of gravity
waves are the main means of dissipation for gravity waves
\changes{\citep{Longuet-Higgins1964, Longuet-Higgins1992, Longuet-Higgins1995,
Sajjadi2002, Melville2015}}. The dissipation of capillary waves has also been
found to increase sharply with increasing steepness of gravity waves, thus delaying the breaking
of gravity waves \changes{\citep{Longuet-Higgins1963, Crapper1970, Deike2015,
Melville2015}}.
Viscosity plays an important role in the energy transfer across scales in
capillary wave turbulence
\citep{Deike2012,Deike2014a,Abdurakhimov2015,Haudin2016} and has also been
suggested to be the main reason for the bi-directional energy cascade observed
in capillary wave turbulence \citep{Abdurakhimov2015}. For instance,
\citet{Deike2014a} showed that the steepness of the power-law spectrum of
capillary wave turbulence increases with increasing viscosity.
\changes{\citet{Perrard2015} studied the propagation of capillary solitons on a
levitated body of liquid and found viscous dissipation to be responsible for a
gradual decrease in wave amplitude, which in turn affects the phase velocity
and leads to a turbulence-like regime at high frequency with respect to the
amplitude power-spectrum.} Capillary waves considerably influence the behavior
and properties, {\em e.g.}~glass transition temperature and effective viscosity, of viscoelastic materials \citep{Monroy1998,Harden1991,Herminghaus2002,Herminghaus2004}, such as gels and
polymers, with critical damping marking the transition between inelastic and
quasielastic behavior \citep{Harden1991,Madsen2004}.

In an inviscid, ideal fluid the angular frequency of capillary waves is given as
\citep{Lamb1932}
\begin{equation}
\omega_{0} = \sqrt{\frac{\sigma k^3}{\tilde{\rho}}} \
,
\label{eq:omegaNull}
\end{equation}
where $\sigma$ is the surface tension coefficient, $k = 2 \pi/\lambda$ is the
wavenumber and $\tilde{\rho} =\rho_\mathrm{a} + \rho_\mathrm{b}$ is the relevant fluid density,
where subscripts $\mathrm{a}$ and $\mathrm{b}$ denote
properties of the two interacting bulk phases.
Thus, the frequency as well as the phase velocity $c_0 = \omega_0/k$ of
capillary waves in inviscid fluids increase with increasing wavenumber, with
$\omega_0 \rightarrow \infty$ for $k \rightarrow \infty$. 

In viscous fluids, viscous stresses attenuate the wave motion and the frequency
takes on the complex form $\omega = \omega_0 + i \Gamma$, with a single
frequency for each wavenumber \citep{Jeng1998}. Three distinct damping regimes
can be identified for capillary waves in viscous fluids (similar to other damped
oscillators):
the underdamped regime for $k < k_\mathrm{c}$, critical damping for $k =
k_\mathrm{c}$ and the overdamped regime for $k > k_\mathrm{c}$.
\changes{At critical damping, with critical wavenumber $k_\mathrm{c}$, the wave
requires the shortest time to return to its equilibrium state without
oscillating, meaning that the real part of the complex angular frequency is
$\mathrm{Re}(\omega) = 0$. Critical damping, hence, represents the transition from the underdamped (oscillatory) regime, with $k < k_\mathrm{c}$ and
$\mathrm{Re}(\omega) > 0$, to the overdamped regime, with $\mathrm{Re}(\omega) =
0$ and $k>k_\mathrm{c}$.} Based on the linearized Navier-Stokes equations, the
dispersion relation of capillary waves in viscous fluids is given as
\citep{Landau1966,Levich1962,Byrne1979,Earnshaw1991}
\begin{equation}
\omega_0^2 + \left(i \omega + 2 \nu k^2\right)^2
 - 4 \nu^2  k^4 \sqrt{1+ \frac{i \omega}{\nu k^2}} = 0\ ,
\label{eq:dispersionRelationFull}
\end{equation}
where $\nu=\mu/\rho$ is the kinematic viscosity and $\mu$ is the dynamic
viscosity.
From Eq.~(\ref{eq:dispersionRelationFull}) the damping rate follows as $\Gamma = 2
\nu k^2$. This damping rate is valid in the weak damping regime, for $k \ll
\sqrt{\omega_0/\nu}$, where viscous damping is considered to be small
\citep{Levich1962,Jeng1998}.

Most research to date has focused on linear wave theory under the assumption of
inviscid fluids or based on the linearized Navier-Stokes equations (weak
damping).
These assumptions have significant limitations with respect to capillary waves
with short wavelength, for which viscous attenuation is a dominant
influence.
The inviscid assumption is only valid for long waves where viscous damping is
negligible, whereas the weak damping assumption is valid when viscous stresses
have a small effect on the dispersion of capillary waves
\citep{Lamb1932,Levich1962,Jeng1998}.
Furthermore, the damping rate $\Gamma$ is not a constant value, as presupposed
by the weak damping assumption, but changes considerably throughout the
underdamped regime \citep{Jeng1998}.
Similarly, the linearized Navier-Stokes equations assume that nonlinear
effects are negligible, which is not the case for short capillary waves or
highly viscous fluids \citep{Behroozi2011}.
As a result, a rational parametrization and consistent characterization that
accurately describes the frequency dispersion of capillary waves in viscous
fluids is not available to date.

The goal of this study is the formulation of a rational
parametrization of the dispersion of capillary waves in viscous fluids, which is valid throughout
the entire underdamped regime and for two-phase systems with arbitrary fluid
properties.
To this end, the dispersion and oscillatory behavior of capillary
waves is studied from a purely hydrodynamic perspective based on continuum
mechanics, assuming that continuum mechanics is valid in the entire underdamped
regime, including critical damping, as previously shown by \citet{Delgado2008}.
The dispersion of capillary waves with small amplitude in different viscous
fluids is computed using an analytical initial-value solution (AIVS) as well as direct
numerical simulation (DNS). 
Given the validity of the underpinning assumptions, AIVS is used
for capillary waves in one-phase systems ({\em i.e.}~a single fluid with a free
surface) as well as two-phase systems in which both phases have the same kinematic viscosity
$\nu$, whereas DNS is applied to extend the study to arbitrary fluid properties. 

A harmonic oscillator model for capillary waves is proposed, which accurately
predicts the wavenumber $k_\mathrm{c}$ at which a capillary wave in arbitrary
viscous fluids is critically damped, and a consistent scaling for capillary
waves is derived from rational arguments. Based on this scaling as well as the
critical wavenumber predicted by the harmonic oscillator model, a 
self-similar characterization of the frequency dispersion of capillary waves in
viscous fluids is introduced. This characterization allows an accurate {\em a
priori} evaluation of the frequency of capillary waves for two-phase systems
with arbitrary fluid properties, and unveils distinct features of capillary
waves that are independent of the fluid properties. Moreover, different methods
to predict the frequency of capillary waves are studied and a new definition of
the effective damping rate is proposed, which provides a more accurate frequency
prediction than commonly used definitions when viscous stresses dominate.

In \changes{Sec.~\ref{sec:characterisation} the characterization of capillary
waves is discussed and Sec.~\ref{sec:computationalMethods} introduces the
applied computational methods}. Section \ref{sec:criticalDamping} examines
critical damping and proposes a harmonic oscillator model to parameterize
capillary waves. Section \ref{sec:propagatingWaves} analyses and discusses the
similarity of the frequency dispersion of capillary waves and
Sec.~\ref{sec:prediction} examines different damping assumptions and frequency
estimates. The article is summarised and conclusions are drawn in
Sec.~\ref{sec:conclusions}.

\changes{\section{Characterization of capillary waves}}
\label{sec:characterisation}
Two physical mechanisms govern the oscillatory motion of capillary waves:
surface tension, which is the dominant dispersive and restoring mechanism, and
viscous stresses in the fluids, which is the prevailing dissipative mechanism
\footnote{Shear and dilatational viscosities of the fluid interface are
neglected in this study.}. Other physical mechanisms, such as inertia or
gradients in surface tension coefficient, can be neglected, since no external
forces ({\em e.g.}~gravity) are imposed on the fluids or the interface, the
fluids are considered pure ({\em i.e.}~free of surfactants), and the fluid
motion induced by the small-amplitude capillary waves is dominated by viscosity
({\em i.e.}~creeping flow).

Capillary (surface tension) effects are quantified by their characteristic
pressure $p_\sigma = {\sigma}/{l}$, where $l$ is a reference lengthscale, and their
characteristic timescale $t_\sigma = \sqrt{{\tilde{\rho} \,
l^3}/{\sigma}}$, which is proportional to the undamped period of a capillary
wave (with $t_\sigma = \omega_0^{-1}$ for $l = k^{-1}$).
With respect to viscous stresses, the characteristic pressure is $p_\mu =
{\tilde{\mu} \, u}/{l}$, where $\tilde{\mu} = \mu_\mathrm{a}+\mu_\mathrm{b}$
and $u$ is a reference velocity. The characteristic timescale associated with
viscous stresses, which is representative of the time required for momentum to
diffuse through a distance $l$, is $t_\mu = {\tilde{\rho} \, l^2}/{\tilde{\mu}}$.
Quantifying the relative importance of surface tension and viscous stresses, the 
characteristic pressures lead to the capillary number
\begin{equation}
\mathrm{Ca} = \frac{p_\mu}{p_\sigma} = \frac{\tilde{\mu} \, u}{\sigma} 
\end{equation}
and the ratio of the characteristic timescales is given by the Ohnesorge
number
\begin{equation}
\mathrm{Oh} = \frac{t_\sigma}{t_\mu} = \frac{\tilde{\mu}}{\sqrt{\sigma \,
\tilde{\rho} \, l}} \ .
\label{eq:ohnesorge}
\end{equation} 
% For the particular case of a capillary wave $\mathrm{Ca} = \mathrm{Oh}$, with $l
% = q^{-1}$ and $u=c_0 = \omega_0/q$.

Assuming that surface tension and viscous stresses are equally important, the
viscocapillary velocity follows from $\mathrm{Ca}=1$ as
\begin{equation}
u_\mathrm{vc} = \frac{\sigma}{\tilde{\mu}} \ ,
\label{eq:viscocapVel}
\end{equation}
and the viscocapillary lengthscale based on $\mathrm{Oh} = 1$ is
\begin{equation}
l_\mathrm{vc} = \frac{\tilde{\mu}^2}{\sigma \, \tilde{\rho}} \ .
\label{eq:viscocapLength}
\end{equation}
The viscocapillary timescale follows from Eqs.~(\ref{eq:viscocapVel}) and
(\ref{eq:viscocapLength}) as
\begin{equation}
t_\mathrm{vc} = \frac{l_\mathrm{vc}}{u_\mathrm{vc}} =
\frac{\tilde{\mu}^3}{\sigma^2 \, \tilde{\rho}} \ .
\label{eq:viscocapTime}
\end{equation}
Note that a similarly defined lengthscale and timescale have been applied in
\citep{Eggers2008} for the long-wave description of the Rayleigh-Plateau
instability on viscous jets.

The main characteristic of an oscillator is its frequency $\omega$, which is
given for a damped oscillator, such as a capillary wave in viscous fluids, as
\begin{equation}
\omega = \omega_0 + i\Gamma = \omega_0 \, \sqrt{1-\zeta^2} \ ,
\label{eq:complexFreq}
\end{equation}
where $\zeta = \Gamma/\omega_0$ is the damping ratio.
\changes{Since capillary waves are dispersive waves, the critical frequency
$k_\mathrm{c}$ at which the wave is critically damped, signified by $\zeta=1$,
is of particular importance as it represents the transition from
underdamped to overdamped behavior.}
\changes{Introducing the dimensionless wavenumber $\hat{k} = k/k_\mathrm{c}$},
the point $(1,1)$ is a uniquely defined point in the $\hat{k}-\zeta$ graph
irrespective of the fluid properties, \changes{since by definition $\zeta=1$ at
$\hat{k} =1$ (see Sec.~\ref{sec:criticalWavenumberValidation} for further
discussion)}.
An accurate estimate of the critical wavenumber $k_\mathrm{c}$ can, thus, serve
as a reference value for the dispersion of an oscillating system.

A second characteristic point of the dispersion of a damped oscillator is the
maximum frequency $\omega_\mathrm{m}$ and the corresponding wavenumber
$k_\mathrm{m}$, as previously also pointed out by \citet{Ingard1988}.
\changes{For capillary waves, this maximum frequency does occur at wavenumbers
noticeably smaller than the critical wavenumber
\citep{Ingard1988,Madsen2004,Hoshino2008}.}
An accurate approximation of the maximum frequency can serve as a reference
value for the frequency and, hence, the oscillatory motion of capillary waves.
\changes{The maximum frequency is also of particular interest for the study and
description of capillary wave turbulence. For freely decaying capillary wave
turbulence an energy transport to higher frequencies beyond the maximum
frequency of freely oscillating capillary waves is physically implausible.}

\section{Computational methods}
\label{sec:computationalMethods}
A single \changes{standing} capillary wave with wavelength $\lambda$ and initial
amplitude $a_0=0.01 \lambda$ is studied in six representative two-phase systems, for
which the fluid properties are given in Table \ref{tab:cases}, using AIVS
(see Sec.~\ref{sec:aivs}) as well as DNS (see Sec.~\ref{sec:dns}). 
\changes{According to the seminal work of \citet{Crapper1957} on progressive capillary
waves, the frequency difference for a progressive capillary wave with
amplitude $a_0 = 0.01 \lambda$ is less than $0.1\%$ compared to the frequency of a capillary
wave with infinitesimal amplitude, which has the same solution for standing and progressive waves.
Although \citet{Crapper1957} neglected viscous stresses in the derivation of the analytical
solution for progressive capillary waves with arbitrary amplitude, this suggests
that there is no appreciable difference between standing and progressive
capillary waves at small amplitude. Focusing on standing capillary waves also
allows to employ the AIVS described in Sec.~\ref{sec:aivs}.}
\begin{table*}[ht]
\begin{center}
\caption{Fluid properties, property ratio $\beta$ [defined in
Eq.~(\ref{eq:propertyRatio})] and solution method(s) of the considered cases.}
\label{tab:cases}
\begin{tabular}{lcccccccc}
Case $ \ $ & $\rho_\mathrm{a} \ [\mathrm{kg}\, \mathrm{m}^{-3}]$  &
$\mu_\mathrm{a} \ [\mathrm{Pa} \, \mathrm{s}]$ & $\rho_\mathrm{b} \ [\mathrm{kg}\,
\mathrm{m}^{-3}]$ & $\mu_\mathrm{b} \ [\mathrm{Pa} \, \mathrm{s}]$ &
$\sigma \ [\mathrm{N} \, \mathrm{m}^{-1}]$ & $\beta$ & AIVS
& DNS\\
\hline
A & $5.0$ & $0.7$ & $5.0$ & $0.7$ & $10^{-3}$ & $6.250 \times 10^{-2}$ & yes &
yes\\ 
B & $2.0$ & $0.01$ & $2000.0$ & $10.0$ & $2.1 \times 10^{-2}$ & $2.495 \times
10^{-4}$ & yes & no \\ 
C & $2.0$ & $0.01$ & $200.0$ & $1.0$ & $2.1 \times 10^{-2}$ & $2.451 \times 10^{-3}$ & yes
& no \\ 
D & $0$ & $0$ & $1000.0$ & $0.001$ & $7.2 \times 10^{-2}$ & $0$ & yes & no \\ 
E & $1.205$ & $1.82 \times 10^{-5}$ & $1000.0$ & $0.001$ & $10^{-5}$ & $7.001
\times 10^{-5}$ & no & yes \\ 
F & $1450.0$ & $2.0$ & $800.0$ & $0.319$ & $7.5 \times 10^{-4}$ & $3.986 \times
10^{-2}$ & no & yes \\ 
\hline
\end{tabular}
\end{center}
\end{table*}

The incompressible flow of isothermal, Newtonian fluids is
governed by the continuity equation
\begin{equation}
\frac{\partial u_i}{\partial x_i} = 0 \label{eq:continuityDNS} 
\end{equation}
and the momentum equations
\begin{equation}
\begin{split}
\frac{\partial  u_i}{\partial t} &+ u_j \frac{\partial
 u_i}{\partial x_j} = - \frac{1}{\rho} \frac{\partial p}{\partial x_i} \\
 &+ \frac{\partial}{\partial x_j} \left[ \nu \left(\frac{\partial
 u_i}{\partial x_j} + \frac{\partial u_j}{\partial x_i} \right)\right] + g_i
 + \frac{f_{\mathrm{\sigma},i}}{\rho} \ ,
\label{eq:momentumDNS}
\end{split}
\end{equation}
often collectively referred to as the Navier-Stokes equations,
where $t$ represents time, $\boldsymbol{u}$ is the flow velocity, $p$ is the
pressure, $\boldsymbol{g}$ is the gravitational acceleration and $\boldsymbol{f}_\sigma$
is the volumetric force due to surface tension acting at the fluid
interface. The hydrodynamic balance of forces acting at the fluid
interface is given as \citep{Levich1969}
\begin{equation}
\begin{split}
\left(p_\mathrm{a} - p_\mathrm{b} \right. &+ \left. \sigma \, \kappa \right)
\hat{m}_i = \left[ \mu_\mathrm{a} \left(\left. \frac{\partial u_i}{\partial
x_j}\right|_\mathrm{a} + \left. \frac{\partial u_j}{\partial x_i}\right|_\mathrm{a} \right)
\right. \\ &- \left. \mu_\mathrm{b} \left( \left. \frac{\partial u_i}{\partial
x_j} \right|_\mathrm{b} + \left. \frac{\partial u_j}{\partial x_i} \right|_\mathrm{b}
 \right) \right] \hat{m}_j - \frac{\partial \sigma}{\partial x_i} \ ,
\label{eq_forceBalanceTheoretical}
\end{split}
\end{equation}
where $\kappa$ is the curvature and
$\boldsymbol{\hat{m}}$ is the unit normal vector (pointing into fluid b) of the fluid interface.

\subsection{Analytical initial-value solution}
\label{sec:aivs}
Based on the linearized Navier-Stokes equations and the interfacial
force-balance given in Eq.~(\ref{eq_forceBalanceTheoretical}),
\citet{Prosperetti1976, Prosperetti1981} analytically derived an
integro-differential equation that provides an exact solution for the
initial-value problem of a capillary wave with small amplitude for the special
cases of a single viscous fluid with a free surface \citep{Prosperetti1976} and
two fluids with equal kinematic viscosity \citep{Prosperetti1981}. Assuming no
gravity and no initial velocity, the amplitude at time $t$ is given as
\citep{Prosperetti1981}
\begin{equation} 
\begin{split}
&a(t) = \frac{4 (1 - 4 \beta_\rho) \, \nu^2 k^4}{8 (1 -4 \beta_\rho) \, \nu^2
k^4 + \omega_0^2} \, a_0 \, \mathrm{erfc}\left(\sqrt{\nu k^2  t}\right) \\
&+ \sum_{i=1}^4 \frac{z_i}{Z_i} \left(\frac{\omega_0^2 \, a_0}{z_i^2-\nu
k^2}\right) \exp\left[\left(z_i^2-\nu k^2\right) t \right] \,
\mathrm{erfc}\left(z_i \sqrt{t}\right)
\label{eq:ampProsp}
\end{split}
\end{equation}
with $a_0$ being the initial amplitude, $Z_1 = (z_2-z_1)(z_3-z_1)(z_4-z_1)$ (and
$Z_2$, $Z_3$ and $Z_4$ calculated by circular permutation of the indices),
$z_1$, $z_2$, $z_3$ and $z_4$ are the roots of the polynomial
\begin{equation}
\begin{split}
z^4 &- 4 \beta_\rho \sqrt{\nu k^2} z^3 + 2 (1-6 \beta_\rho) \nu k^2 z^2 
\\ &+ 4 (1 - 3 \beta_\rho) (\nu k^2)^{3/2} z + (1 - 4 \beta_\rho) \nu^2 k^4 +
\omega_0 = 0
\label{eq:ampProsp2}
\end{split}
\end{equation}
and $\beta_\rho = \rho_\mathrm{a} \rho_\mathrm{b}/\tilde{\rho}^2$.
Equation (\ref{eq:ampProsp}) is solved at time intervals $\Delta t = (200 \,
\omega_0)^{-1}$, {\em i.e.}~with $200$ solutions per undamped period, which
provides a sufficient temporal resolution of the evolution of the capillary
wave.

\subsection{DNS methodology}
\label{sec:dns}
DNS of the entire two-phase system, including both bulk phases as well as the
fluid interface, are conducted by resolving all relevant scales in space and
time. The governing equations are solved numerically using a coupled finite-volume
framework with collocated variable arrangement \citep{Denner2014}.
The continuity equation, Eq.~(\ref{eq:continuityDNS}), is discretized using a
balanced-force implementation of the momentum-weighted interpolation method
\citep{Denner2014}, which couples pressure and velocity.
The momentum equations, given in Eq.\ (\ref{eq:momentumDNS}), are discretized
using second-order accurate schemes in space and time, as detailed in
\citep{DennerThesis2013}.

The Volume-of-Fluid (VOF) method \citep{Hirt1981} is adopted to describe
the interface between the immiscible bulk phases.
The local volume fraction of both phases is represented by the
color function $\gamma$, defined as $\gamma = 0$ in fluid $\mathrm{a}$ and
$\gamma = 1$ in fluid $\mathrm{b}$, with the interface located in regions with a
color function value of $0 < \gamma < 1$. The local density $\rho$ and
dynamic viscosity $\mu$ are calculated using an arithmetic average based on the
color function $\gamma$. The colour function $\gamma$ is advected by the linear
advection equation
\begin{equation}
\frac{\partial \gamma}{\partial t} + u_i \frac{\partial \gamma}{\partial x_i}
= 0 \ ,
\label{eqn_vofAdvection}
\end{equation}
which is discretized using a compressive VOF method
\citep{Denner2014d,Denner2014a}.

Surface tension is discretized using the continuum surface force (CSF) model
\citep{Brackbill1992} as a volume force acting in the interface region
\begin{equation}
\boldsymbol{f}_{s} = \sigma \, \kappa \, \nabla \gamma \ .
\label{eq_csfSurfaceTensionGradient}
\end{equation}
The interface curvature is computed as $\kappa = h_{xx}/(1+h_x^2)^{3/2}$, where
$h_x$ and $h_{xx}$ represent the first and second derivatives with respect to
the $x$-axis of height $h$ of the color function $\gamma$ in the direction
normal to the interface, calculated by means of central differences.
No convolution is applied to smooth the surface force or the color function
field \citep{Denner2013}.
 
The applied two-dimensional computational domain has the dimensions
$\lambda \times 3\lambda$, all boundaries of which are treated as free-slip walls, and is
represented by an equidistant Cartesian mesh with mesh spacing $\Delta x =
\lambda/100$. The initial amplitude of the capillary wave is $a_0 = 0.01
\lambda$, no gravity is acting and the flow field is initially stationary.
The time-step applied to solve the governing equations is
$\Delta t = (200 \, \omega_0)^{-1}$, which allows a direct comparison with
the AIVS results, fulfils the capillary time-step constraint \citep{Denner2015}
and results in a Courant number of $\mathrm{Co} = \Delta t \,
|\boldsymbol{u}|/\Delta x < 10^{-2}$.
 
\subsection{Validation of the DNS methodology}
\begin{figure}
  \centerline{\includegraphics[width=0.45\textwidth]{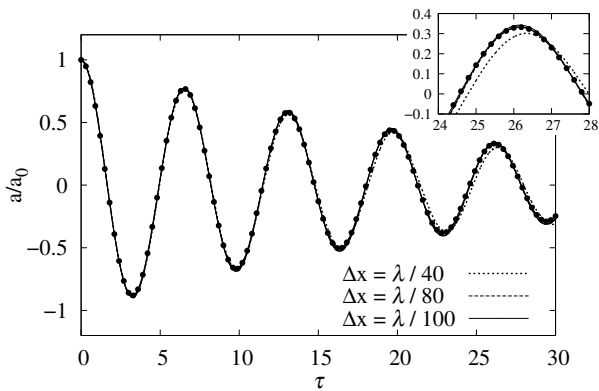}}
  \caption{DNS results of the amplitude of a capillary wave of Case A with
  wavenumber $k = 10^{-4} k_\mathrm{c}$ as a function of dimensionless time
  $\tau = t \omega_0$ for different mesh spacings $\Delta x$.
  The analytical solution according to Eq.~(\ref{eq:ampProsp}) is shown as
  solid circles. The inset shows the amplitude for a particular time interval.}
\label{fig:mfMeshResolution}
\end{figure}

The DNS methodology is validated against AIVS using Case
A as a representative case. In order to confirm that the solution is
mesh-independent for the chosen mesh resolution of $\Delta x =\lambda/100$, the transient evolution of the wave
amplitude obtained with three different mesh resolutions $\Delta x \in
\{\lambda/40, \lambda/80, \lambda/100\}$ is compared for Case A with wavenumber
$k=10^{-4} k_\mathrm{c}$, with $k_\mathrm{c}$ being the critical wavenumber
(discussed in detail in Sec.~\ref{sec:criticalDamping}).
Figure \ref{fig:mfMeshResolution} shows that the DNS accurately predicts the
transient evolution of the wave amplitude with $\Delta x = \lambda/80$ and
$\Delta x = \lambda/ 100$, exhibiting only
minor differences between the results obtained on both meshes, and also compared
to the results obtained with AIVS. The DNS result obtained on a mesh with
$\Delta x = \lambda/40$, however, shows a visible and continuously growing
error in frequency as time progresses.

\begin{figure}
  \centerline{\includegraphics[width=0.45\textwidth]{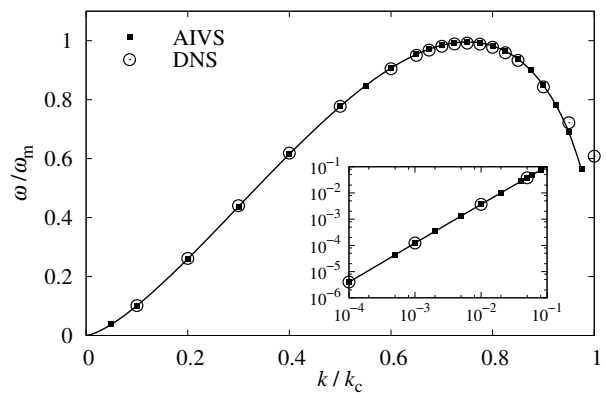}}
  \caption{Comparison of the dimensionless frequency $\omega/\omega_\mathrm{m}$
  as a function of the dimensionless wavenumber $k/k_\mathrm{c}$ obtained with
  AIVS and DNS for Case A. The inset shows the frequency at small wavenumbers and the solid
  line represents a spline-fit of the AIVS result.}
\label{fig:comparisonAIVSDNS}
\end{figure}

Figure \ref{fig:comparisonAIVSDNS} shows the dimensionless frequency
$\omega/\omega_\mathrm{m}$, where $\omega_\mathrm{m}$ is the maximum frequency,
as a function of dimensionless wavenumber $k/k_\mathrm{c}$ obtained with AIVS
and DNS.
The frequency $\omega$ of the capillary waves is calculated directly
from the transient evolution of the amplitude, see for instance
Fig.~\ref{fig:mfMeshResolution}, as $\omega = \pi/t_1$, where $t_1$ is the time
of the first extrema of the wave amplitude ({\em e.g.}~the minima of $a$ in
Fig.~\ref{fig:mfMeshResolution} at $\tau \approx \pi$).
The comparison shown in Fig.~\ref{fig:comparisonAIVSDNS} suggests that the DNS
results are accurate for $k \leq 0.9 \, k_\mathrm{c}$. For $k > 0.9 \,
k_\mathrm{c}$ the magnitude of the amplitude at $t_1$ becomes $a \sim 10^{-6}
\lambda$, which is comparable to the residuals of the numerical solution
procedure as well as the model and discretization errors of the DNS methodology.
It can therefore be concluded that the DNS is in very good agreement with the
AIVS and is a suitable tool to study capillary waves with $k\leq 0.9 \,
k_\mathrm{c}$.

\section{Critical damping}
\label{sec:criticalDamping}
As explained in Sec.~\ref{sec:characterisation}, {\em a priori} knowledge of the
critical wavenumber is important for a consistent characterization 
of capillary waves. Based on the dispersion relation \changes{following from
the linearized Navier-Stokes equations}, given in
Eq.~(\ref{eq:dispersionRelationFull}), \citet{Byrne1979} and \citet{Hoshino2008} derived the critical wavenumber as
\begin{equation}
k_\mathrm{c}^\mathrm{B} = \frac{1.725}{l_\mathrm{vc}} \ ,
\label{eq:criticalWavenumberHoshino}
\end{equation}
where superscript $\mathrm{B}$ is reference to the original author(s),
since for $k \geq k_\mathrm{c}^\mathrm{B}$ the roots of
Eq.~(\ref{eq:dispersionRelationFull}) are purely imaginary. 
This was confirmed experimentally for capillary waves on water-air and
glycerol-air interfaces
\citep{Byrne1979}. A similar critical wavenumber $k_\mathrm{c}^\mathrm{K}
= 1.7214 \, l_\mathrm{vc}^{-1}$ has been proposed by \citet{Katyl1968}
based on light scattering experiments.
In what follows, a new definition of the critical wavenumber is derived based
on a harmonic oscillator model, which is shown to be in excellent agreement with
AIVS and DNS results.

\subsection{Harmonic oscillator model}
\label{sec:hom}
Assuming the surface tension coefficient and the viscosity
are constant, the oscillation of a capillary wave with small amplitude is
described as a harmonic oscillator.
For a mass oscillating harmonically in a viscous fluid, the displacement $s$ is
given by the second-order ordinary differential equation \citep{Landau1969}
\begin{equation}
\ddot{s} + 2 \, \zeta \, \omega_0 \, \dot{s} + \omega_0^2 \, s = 0 \ .
\end{equation}
The undamped frequency $\omega_0 = \sqrt{c/m}$ is a function of the spring
constant $c$ and the mass $m$ of the oscillator, while the damping ratio is
\begin{equation}
\zeta = \frac{b}{2 \sqrt{m c}} \ ,
\label{eq:dampingRatioGeneral}
\end{equation}
with $b$ being the viscous damping coefficient. With respect to capillary waves,
the spring constant is $c = \sigma$ and the mass is $m =
\tilde{\rho}/k^3$, so that $\omega_0 = \sqrt{c/m} = \sqrt{\sigma
k^3/\tilde{\rho}}$, see Eq.~(\ref{eq:omegaNull}).
Following a dimensional analysis, the viscous damping coefficient can be
defined as $b = \tilde{\mu} L$, where $L$ is the damping length.

Based on the lengthscale $l_\nu = \sqrt{2\nu/\omega_0}$ for the
penetration depth of the vorticity generated by a capillary wave
\citep{Landau1966,Jenkins1997b}, the critical damping length is
defined as
\begin{equation}
L_\mathrm{c} = \sqrt{\frac{2 \tilde{\mu}}{ \omega_{0,\mathrm{c}} \,
\tilde{\rho}}} \ ,
\end{equation}
where $\omega_{0,\mathrm{c}} = \sqrt{\sigma k_\mathrm{c}^3/\tilde{\rho}}$ is the
undamped frequency at the critical wavenumber $k_\mathrm{c}$.
Inserting the critical viscous damping coefficient $b_\mathrm{c} = \tilde{\mu}
L_\mathrm{c}$, the mass at critical damping $m_\mathrm{c} =
\tilde{\rho}/k_\mathrm{c}^3$ and the spring constant $c=\sigma$ into
Eq.~(\ref{eq:dampingRatioGeneral}), the critical damping ratio becomes
\begin{equation}
\zeta_\mathrm{c} = \frac{b_\mathrm{c}}{2 \sqrt{m_\mathrm{c} c}} =
\frac{\tilde{\mu} L_\mathrm{c}}{2 \sqrt{\sigma \tilde{\rho} / k_\mathrm{c}^3}} =
1 \ .
\label{eq:dampingRatioCritical}
\end{equation}
From Eq.~(\ref{eq:dampingRatioCritical}) the critical wavenumber based on
the harmonic oscillator model (indicated by superscript h.o.) readily follows as
\begin{equation}
k_\mathrm{c}^{\ast,\mathrm{h.o.}} = 2^{2/3} \, \frac{\sigma 
\tilde{\rho}}{\tilde{\mu}^2} = \frac{2^{2/3}}{l_\mathrm{vc}} 
% \approx \frac{1.587}{l_\mathrm{vc}}
 \ ,
\label{eq:criticalWavenumberEqual}
\end{equation}
where superscript $\ast$ marks it as a newly proposed value (this notation is
applied consistently throughout the manuscript).

However, the critical wavenumber given by Eq.~(\ref{eq:criticalWavenumberEqual})
is only accurate for bulk phases with equal density and viscosity.
For two-phase systems where the bulk phases have different properties, the
critical wavenumber is expanded by the dimensionless property ratio
\begin{equation}
\beta = \frac{\rho_\mathrm{a}
\rho_\mathrm{b}}{\tilde{\rho}^2} \,
\frac{\nu_\mathrm{a} \nu_\mathrm{b}}{\tilde{\nu}^2}  
%=\frac{\mu_\mathrm{a} \mu_\mathrm{b}}{\tilde{\rho}^2 \, \tilde{\nu}^2} 
\ ,
\label{eq:propertyRatio}
\end{equation}
with $\tilde{\nu} = \nu_\mathrm{a}+\nu_\mathrm{b}$, to become 
\begin{equation}
k^\ast_\mathrm{c} = k_\mathrm{c}^{\ast,\mathrm{h.o.}} \,  (1.0625 -
\beta) = \frac{2^{2/3}}{l_\mathrm{vc}} \, (1.0625 - \beta) \ .
\label{eq:criticalWavenumber}
\end{equation}
This equation accurately predicts the critical wavenumber for capillary waves in
fluids with arbitrary properties, as shown below. The property ratio $\beta$ is
bounded by the case of a single fluid with a free surface ({\em i.e.}
$\nu_\mathrm{b}=\rho_\mathrm{b} =0$) for which $\beta = 0$ and the two-phase
case with $\rho_\mathrm{a}=\rho_\mathrm{b}$ and $\nu_\mathrm{a}=\nu_\mathrm{b}$
for which $\beta = 0.0625$.
Note that the density ratio in Eq.~(\ref{eq:propertyRatio}) is also included in
the AIVS as proposed by \citet{Prosperetti1981}, see Eqs.~(\ref{eq:ampProsp})
and (\ref{eq:ampProsp2}), where it is associated with the continuity of
tangential stresses at the interface, and appears in the damping rate proposed
by \citet{Jeng1998} for small damping.

\changes{Comparing the critical wavenumber $k_\mathrm{c}^\ast$ predicted by the proposed
harmonic oscillator model, Eq.~(\ref{eq:criticalWavenumber}), with the critical
wavenumber $k_\mathrm{c}^\mathrm{B}$ derived from the linearized Navier-Stokes
equations (weak damping assumption), Eq.~(\ref{eq:criticalWavenumberHoshino}),
yields a difference ranging from $2.3\%$ for $\beta = 0$ to $8.7 \%$ for $\beta
= 0.0625$. Hence, the difference is largest for two-phase systems in which both
bulk phases have equal properties ($\beta = 0.0625$) and smallest for a single
fluid with a free surface ($\beta = 0$).}
 
\changes{In practice this means, for instance, that a capillary wave on a
water-air interface at room temperature has a critical wavenumber of
$k_\mathrm{c}^\ast=1.17 \times 10^{8} \, \mathrm{m}^{-1}$, which corresponds to
a critical wavelength of $\lambda_\mathrm{c}^\ast = 5.37 \times 10^{-8} \,
\mathrm{m}$.
This compares to a critical wavenumber of $k_\mathrm{c}^\mathrm{B} = 1.24 \times
10^{8} \, \mathrm{m}^{-1}$ ($\lambda_\mathrm{c}^\mathrm{B} = 5.07 \times 10^{-8}
\, \mathrm{m}$) based on the linearized Navier-Stokes equations, a difference
in wavenumber of $6.95 \times 10^6 \, \mathrm{m}^{-1}$, which corresponds
to a difference in critical wavelength of $3.03
\times 10^{-9} \, \mathrm{m}$. For a capillary wave on an interface between
glycerol and air, the critical wavenumber is $k_\mathrm{c}^\ast=60 \,
\mathrm{m}^{-1}$, corresponding to a critical wavelength of
$\lambda_\mathrm{c}^\ast = 0.11 \, \mathrm{m}$, which exemplifies the wide range
of critical wavenumbers found in typical engineering applications and natural processes, from the
nanoscale to the macroscale.}

\subsection{Balance of scales}
\label{sec:ohnesorge}
\changes{It is worth recalling, that surface tension and viscous stresses are the
dominant physical mechanisms for capillary waves, as discussed in
Sec.~\ref{sec:characterisation}.}
The Ohnesorge number, see Eq.~(\ref{eq:ohnesorge}), which compares the characteristic
timescales of surface tension and viscous stresses, at critical damping with $l =
1/k_\mathrm{c}^{\ast}$ is
\begin{equation}
\mathrm{Oh}_{\mathrm{c}} = \tilde{\mu} \,
\sqrt{\frac{k_\mathrm{c}^\ast}{\sigma \tilde{\rho}}} = 2^{1/3} \, \sqrt{1.0625-\beta} \ .
\label{eq:criticalOhnesorge}
\end{equation}
Rearranging this equation for $k_\mathrm{c}^\ast$ leads back to
Eq.~(\ref{eq:criticalWavenumber}).
This suggests that at critical damping the dispersive/restoring mechanism
(surface tension) and the dissipative mechanism (viscous stresses) reach a
specific balance; hence, occur at a specific relative timescale.
\changes{Since, with lengthscale $l=1/k_\mathrm{c}^\ast$, the capillary
timescale is the inverse of the frequency of a capillary wave with wavenumber
$k_\mathrm{c}^\ast$ in inviscid fluids, $t_\sigma = \omega_{0,\mathrm{c}}^{-1} =
\sqrt{\sigma k_\mathrm{c}^\ast/\tilde{\rho}}$, the balance of capillary and
viscous effects described by Eq.~(\ref{eq:criticalOhnesorge}) means that the
viscous timescale at critical damping is $t_\mu \sim \beta^{-1/2} \,
\omega_{0,\mathrm{c}}^{-1}$.
This suggests that the viscous attenuation of capillary waves is only dependent
on the frequency $\omega_0$ (and, consequently, the phase velocity
$c_0=\omega_0/k$) and on the fluid property ratio $\beta$.}

\subsection{Validation}
\label{sec:criticalWavenumberValidation}
\begin{figure}
\subfloat[Scaled based on $k_\mathrm{c}^\ast$]{
\includegraphics[width=0.225\textwidth]{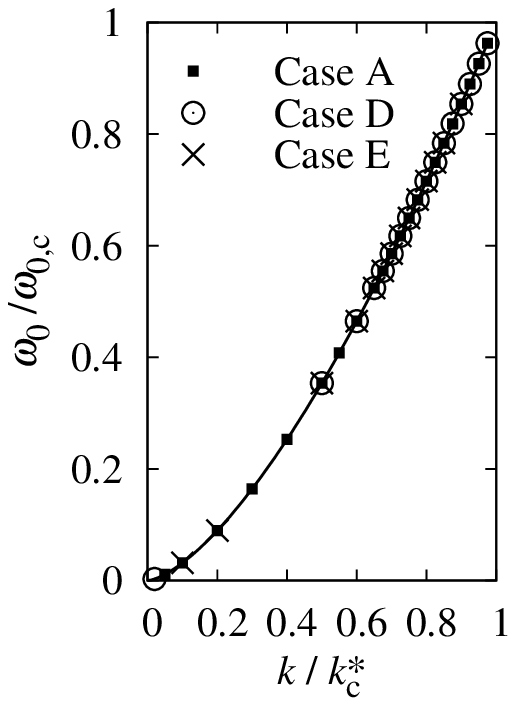}}
\subfloat[Scaled based on $k_\mathrm{c}^\mathrm{B}$]{
\includegraphics[width=0.225\textwidth]{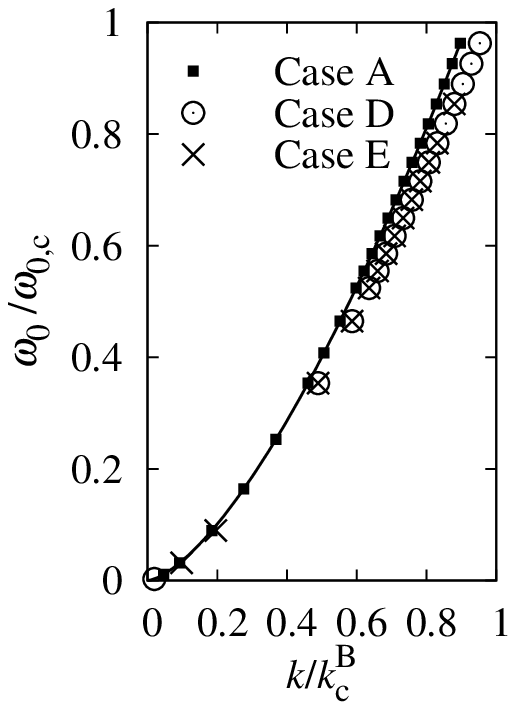}}
  \caption{Dimensionless undamped frequency
  $\hat{\omega}_0=\omega_0/\omega_{0,\mathrm{c}}$ as a function of
  dimensionless wavenumber $k/k_\mathrm{c}^\ast$, based on the proposed critical
  wavenumber given in Eq.~(\ref{eq:criticalWavenumber}), and
  $k/k_\mathrm{c}^\mathrm{B}$, based on the critical wavenumber given in
  Eq.~(\ref{eq:criticalWavenumberHoshino}).}
\label{fig:baseValueScaling}
\end{figure}

Figure \ref{fig:baseValueScaling} shows the dimensionless undamped frequency
$\omega_0/\omega_{0,\mathrm{c}}$ as a function of the dimensionless wavenumbers
$k/k_\mathrm{c}^\ast$, proposed in Eq.~(\ref{eq:criticalWavenumber}), and
$k/k_\mathrm{c}^\mathrm{B}$, given by Eq.~(\ref{eq:criticalWavenumberHoshino}),
with $\omega_{0,\mathrm{c}}$ being the undamped frequency at $k_\mathrm{c}$.
The dimensionless frequencies for the displayed cases fall on a single line when
the proposed critical wavenumber $k^\ast_\mathrm{c}$ is used as a basis for the
normalization, suggesting that $k_\mathrm{c}^\ast$ is a characteristic value
of the frequency dispersion of capillary waves.
In contrast, no consistent correlation between $\omega_0/\omega_{0,\mathrm{c}}$
and $k/k_\mathrm{c}^\mathrm{B}$ is observed, see
Fig.~\ref{fig:baseValueScaling}b.
This supports the findings of \citet{Jeng1998}, who reported that weak damping
is not an adequate assumption for the entire underdamped regime, since
$k_\mathrm{c}^\mathrm{B}$ is derived from the dispersion relation given in
Eq.~(\ref{eq:dispersionRelationFull}) and is, therefore, based on the weak
damping assumption.

\begin{figure}[t]
\begin{center}
\changesFig{\includegraphics[height=0.3\textwidth]{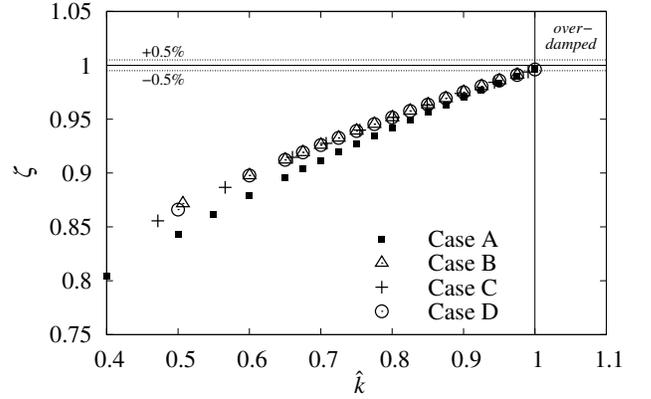}}
\caption{Damping ratio $\zeta$ as a function of dimensionless wavenumber
$\hat{k} = k/k_\mathrm{c}^\ast$ for the considered cases obtained with AIVS,
where point $(1,1)$ represents critical damping. 
\changes{Note that $\zeta \geq 1$ for $\hat{k}<1$ and $\zeta \leq 1$ for
$\hat{k}>1$ are not part of the physical parameter space.}}
\label{fig:dampingRatioAIVS}
\end{center}
\end{figure}

Figure \ref{fig:dampingRatioAIVS} shows the damping ratio $\zeta$ as a function
of the dimensionless wavenumber $\hat{k}=k/k_\mathrm{c}^\ast$ for the cases
considered with AIVS.
Despite the large variety of fluid properties of these cases ({\em
i.e.}~spanning the entire possible $\beta$-range), the solutions consistently
approach and cross $\zeta = 1$ at critical damping ($\hat{k} = 1$) within the expected
margins of error. Note that for $\hat{k}=1$ the AIVS still shows oscillatory
behavior of the wave amplitude. However, the wave amplitude at the first
extrema has a magnitude of $|a_1| \approx 10^{-14}$.
This error can be attributed to the finite precision of floating point
arithmetic, which is approximately $2.22 \times 10^{-16}$ on a 64-bit system
according to IEEE Standard 754, in conjunction with the large number of
conducted time-steps ({\em e.g.}~$6996$ time-steps to $t_1$ for Case A).
Moreover, the waves exhibit no oscillations for $\hat{k} = 1.01$.

\section{Dispersion similarity}
\label{sec:propagatingWaves}
Having the means to accurately predict the critical wavenumber, and considering
that a universal formulation of the critical wavenumber is possible, raises the
question if the dispersion of capillary waves is in fact self-similar following
an appropriate scaling. 

\begin{figure}
\begin{center}
  \subfloat[AIVS results]
  {\includegraphics[width=0.45\textwidth]{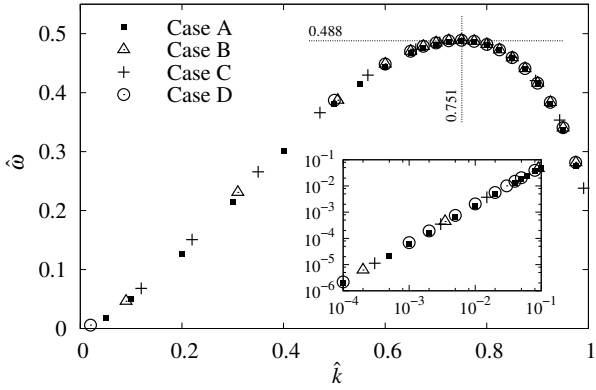}
  \label{fig:dispersionAIVS}}\\
  \subfloat[DNS results]
  {\includegraphics[width=0.45\textwidth]{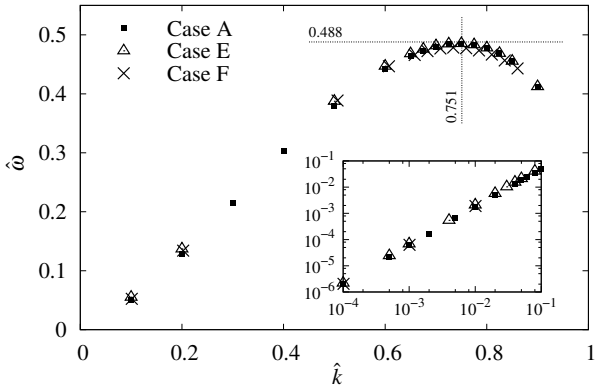}
  \label{fig:dispersionDNS}} 
  \caption{Dimensionless frequency $\hat{\omega}=\omega \,
  t_\mathrm{vc}$ as a function of dimensionless wavenumber
  $\hat{k}=k/k_\mathrm{c}^\ast$ obtained with AIVS and DNS. The inset shows the
  frequency of capillary waves with small wavenumbers.}
\label{fig:dispersionSelfSim}
\end{center}
\end{figure}

Figure \ref{fig:dispersionAIVS} shows the dimensionless frequency $\hat{\omega}
= \omega \, t_\mathrm{vc}$ as a function of
dimensionless wavenumber $\hat{k}=k/k_\mathrm{c}^\ast$ obtained with AIVS for
Cases A-D. The results fall on a single line in the $\hat{k}-\hat{\omega}$
graph throughout the entire underdamped regime, with a single dimensionless
frequency $\hat{\omega}$ for every dimensionless wavenumber $\hat{k}$. The same
similarity is exhibited by the DNS results of Cases A, E and F, shown in
Fig.~\ref{fig:dispersionDNS}. Thus, normalizing the frequency based on the
viscocapillary timescale $t_\mathrm{vc}$, see Eq.~(\ref{eq:viscocapTime}), and
normalizing the wavenumber with the proposed critical wavenumber
$k_\mathrm{c}^\ast$, see Eq.~(\ref{eq:criticalWavenumber}), leads to a
self-similar characterization of the frequency dispersion of capillary waves.
Since a specific dimensionless frequency $\hat{\omega}$ can be
associated with each dimensionless wavenumber $\hat{k}$, the frequency for
capillary waves of any wavenumber is readily available based on the solution
for one particular two-phase system, for instance obtained with AIVS or from
experiments, and irrespective of the fluid properties. 
\changes{This similarity also provides further evidence for the balance
of capillary and viscous timescales and the ensuing dependency of the
viscous attenuation on fluid properties and wavenumber only, as
proposed in Sec.~\ref{sec:ohnesorge}.}

Based on the presented AIVS and DNS results the maximum frequency is
\begin{equation}
\omega_\mathrm{m}^\ast = \frac{0.488}{t_\mathrm{vc}}  \ ,
\end{equation}
which accurately quantifies the maximum frequency in all considered cases.
Furthermore, the maximum frequency is consistently observed in the presented
results at approximately
\begin{equation}
k^\ast_\mathrm{m} = 0.751 \, k_\mathrm{c}^\ast \ .
\end{equation} 
Note that the correction using the property ratio $\beta$ included in
$k_\mathrm{c}^\ast$, see Eq.~(\ref{eq:criticalWavenumber}), also applies to
$k^\ast_\mathrm{m}$. Thus, the maximum frequency is a characteristic value of
the dispersion of capillary waves, as discussed in Sec.~\ref{sec:characterisation}.
\citet{Ingard1988} proposed empirical estimates for the maximum frequency
$\omega_\mathrm{m}^\mathrm{I} = 0.46244 \, t_\mathrm{vc}^{-1}$ and the
corresponding wavenumber $k_\mathrm{m}^\mathrm{I} = 1.2797 \,
l_\mathrm{vc}^{-1}$, which however result in a smaller maximum frequency and
larger corresponding wavenumber than observed in the results presented in
Fig.~\ref{fig:dispersionSelfSim}.

\changes{ The maximum frequency of a capillary wave on a water-air interface is
$\omega_\mathrm{m}^\ast = 2.52 \times 10^9 \, \mathrm{s}^{-1}$ at
$k_\mathrm{m}^\ast=8.79 \times 10^{7} \, \mathrm{m}^{-1}$
($\lambda_\mathrm{m}^\ast = 7.15 \times 10^{-8} \, \mathrm{m}$), whereas for a
capillary wave on an interface between glycerol and air, the maximum frequency
of $\omega_\mathrm{m}^\ast = 0.9 \, \mathrm{s}^{-1}$ occurs at
$k_\mathrm{m}^\ast=45.1 \, \mathrm{m}^{-1}$ ($\lambda_\mathrm{m}^\ast = 0.14 \,
\mathrm{m}$). The range of maximum frequencies for practically relevant two-phase systems, hence, spans
over several orders of magnitude, similar to the range of practically relevant
critical wavenumbers discussed in Sec.~\ref{sec:hom}.
The maximum frequency for an water-air system ($\omega_\mathrm{m} \approx  10^9
\, \mathrm{s}^{-1}$) is several orders of magnitude larger than the typical
maximum frequencies ($\omega \lesssim 10^3-10^4 \, \mathrm{s}^{-1}$) with
appreciable energy \citep{Deike2012,Deike2014,Abdurakhimov2015} of freely
decaying capillary wave turbulence and should, thus, not be a
limiting factor for practical applications or experimental studies of capillary
wave turbulence.}

\section{Damping regimes and assumptions}
\label{sec:prediction}
\begin{figure}[t]
  \centerline{\includegraphics[width=0.45\textwidth]{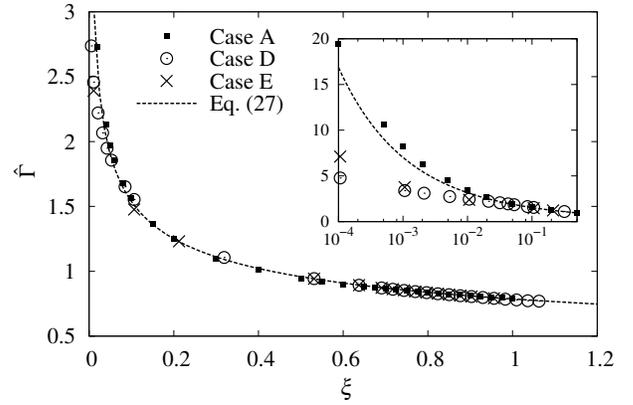}}
  \caption{Dimensionless damping rate $\hat{\Gamma}=\Gamma \tilde{\rho}
  /(\tilde{\mu} k^2)$ as a function of dimensionless wavenumber $\xi =
  k/k_\mathrm{c}^{\ast,\mathrm{h.o.}}$ obtained with AIVS for Cases A ($\beta
  =0.0625$) and D ($\beta = 0$) and with DNS for Case E ($\beta = 7 \times
  10^{-5}$).
  The inset shows the dimensionless damping rate at small wavenumbers.}
\label{fig:gammaNorm}
\end{figure}

As mentioned in the introduction, the damping provided by viscous stresses in
the underdamped regime is not accurately described by commonly used assumptions.
Figure \ref{fig:gammaNorm} shows the dimensionless damping rate
$\hat{\Gamma}=\Gamma \tilde{\rho} /(\tilde{\mu} k^2)$ as a function of
dimensionless wavenumber $\xi = k/k^{\ast,\mathrm{h.o.}}_\mathrm{c}$, which is
based on the critical wavenumber obtained from the proposed harmonic oscillator
model $k^{\ast,\mathrm{h.o.}}_\mathrm{c}$, see
Eq.~(\ref{eq:criticalWavenumberEqual}), without the expansion using the property
ratio $\beta$. The dimensionless damping rate $\hat{\Gamma}$ changes
considerably for different $\xi$, and at small wavenumbers $\hat{\Gamma}$ is
also dependent on the fluid properties. A similar observation can be made in
Fig.~\ref{fig:dampingRatioAIVS}, since the damping ratio $\zeta$ develops
differently as a function of $\hat{k}$ for each of the considered two-phase
systems and, thus, the capillary waves experience a different effective damping
rate $\Gamma = \zeta \omega_0$ in each two-phase system.

For $\xi \gtrsim 0.1$, the results shown in Fig.~\ref{fig:gammaNorm} suggest
a similarity of $\hat{\Gamma}$, since all results fall on a single line, thus
delineating a distinct regime of the dispersion of capillary waves.
In this regime of significant damping the dimensionless damping rate is well approximated by the correlation
\begin{equation}
\hat{\Gamma} = 0.6868 \, \xi^{-1/4} + 0.1 \, \xi^{-1/2} \ ,
\label{eq:dampingFit}
\end{equation}
as seen in Fig.~\ref{fig:gammaNorm}. Hence, the frequency in this regime can be readily estimated as
\begin{equation} \bar{\omega}^\ast =
\omega_0 + i \frac{\tilde{\mu} k^2}{\tilde{\rho}} \, (0.6868 \, \xi^{-1/4} +
0.1 \, \xi^{-1/2}) \ .
\label{eq:freqStrongDamping}
\end{equation}

\begin{figure*}[ht]
\begin{center}
\changesFig{ \subfloat[Error $\varepsilon$ based on $\bar{\omega}^\ast$]
  {\includegraphics[width=0.275\textwidth]{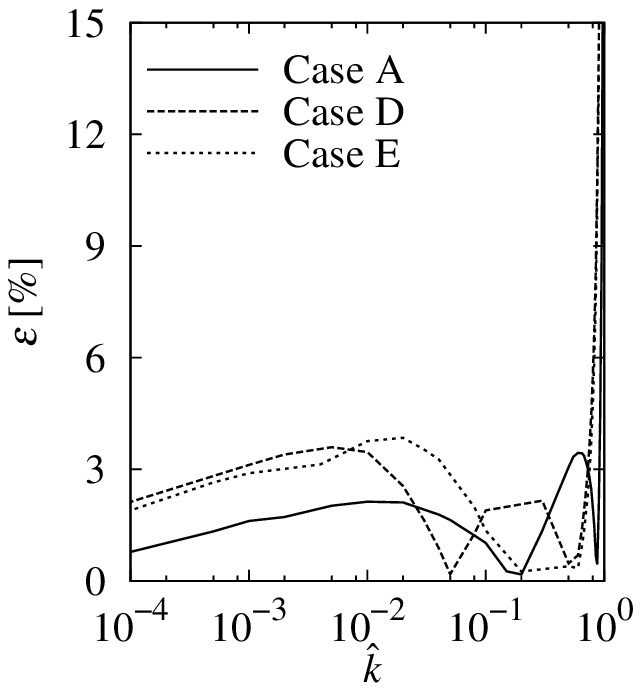}
  \label{fig:errorStrong}} 
  \subfloat[Error $\varepsilon$ based on $\bar{\omega}^\mathrm{(i)}$]
  {\includegraphics[width=0.275\textwidth]{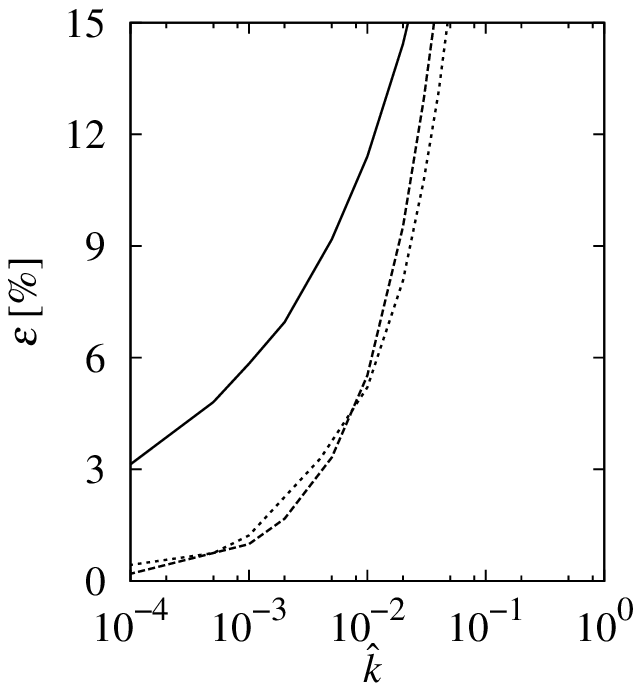}
  \label{fig:errorInviscid}} 
  \subfloat[Error $\varepsilon$ based on $\bar{\omega}^\mathrm{(ii)}$]
  {\includegraphics[width=0.275\textwidth]{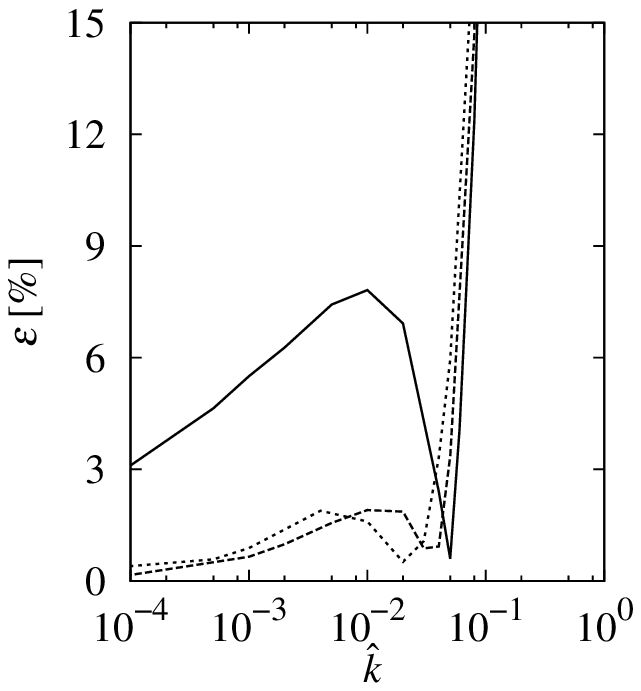}
  \label{fig:errorWeak}} }
  \caption{Relative error $\varepsilon$, see Eq.~(\ref{eq:dispersionError}), of
  the frequency predictions given in Eqs.~(\ref{eq:freqStrongDamping}),
  (\ref{eq:freqInviscid}) and (\ref{eq:freqWeakDamping}) with respect to the
  frequencies obtained from AIVS and DNS as a function of dimensionless wavenumber $\hat{k}$.}
\label{fig:errorDispersion}
\end{center}
\end{figure*}

In order to show the differences between frequency estimates based on different
assumptions, Fig.~\ref{fig:errorDispersion} shows the relative error
\begin{equation}
\varepsilon(\hat{k}) = \frac{|\bar{\omega}(\hat{k})-\omega(\hat{k})|}{{\omega}(\hat{k})} \ .
\label{eq:dispersionError}
\end{equation}
of $\bar{\omega}^\ast$ alongside frequency estimates given by the undamped
frequency
\begin{equation}
\bar{\omega}^\mathrm{(i)} = \omega_0  \ ,
\label{eq:freqInviscid}
\end{equation}
with $\omega_0$ defined in Eq.~(\ref{eq:omegaNull}), and based on the weak
damping assumption
\begin{equation}
\bar{\omega}^\mathrm{(ii)} = \omega_0 + i \Gamma_0 \ ,
\label{eq:freqWeakDamping}
\end{equation}
with $\Gamma_0 = 2 \tilde{\mu} k^2/\tilde{\rho}$ \citep{Jeng1998}.
The frequency estimate proposed in Eq.~(\ref{eq:freqStrongDamping}) exhibits a
high accuracy throughout the underdamped regime, as seen in
Fig.~\ref{fig:errorStrong}. Due to the
definition of the relative error given in Eq.~(\ref{eq:dispersionError}),
$\varepsilon$ increases rapidly as $\hat{k} \rightarrow 1$ since $\omega
\rightarrow 0$. At small wavenumbers (long wavelength) the frequency of 
capillary waves in viscous fluids is well approximated by neglecting the
influence of viscous stresses using the frequency estimate
$\bar{\omega}^\mathrm{(i)}$, as seen in Fig.~\ref{fig:errorInviscid}, with
$\varepsilon < 6\%$ for $\hat{k}\leq10^{-3}$, and a rapidly reducing error for
decreasing wavenumbers. The frequency estimate $\bar{\omega}^\mathrm{(ii)}$
based on the weak damping assumption provides a good estimate up to $\hat{k}
\approx 0.07$, with $\varepsilon < 8\%$, as observed in
Fig.~\ref{fig:errorWeak}.

Note that the error of $\bar{\omega}^\mathrm{(i)}$ and
$\bar{\omega}^\mathrm{(ii)}$ is substantially smaller for Cases D and E,
{\em i.e.}~for two-phase systems with very different fluid properties of the
bulk phases ($\beta \rightarrow 0$), than for Case A ($\beta=0.0625$). This is
presumably the reason why linear wave theory based on inviscid fluids or the weak damping
assumption is frequently found to be in very good agreement with experiments and numerical simulations that
have so far mostly focused on gas-liquid systems ({\em e.g.}~water-air systems),
for which $\beta \approx 0$. For example, \citet{Byrne1979} reported the critical
wavenumber $k_\mathrm{c}^\mathrm{B}$, given in Eq.~(\ref{eq:criticalWavenumberHoshino}), which is
derived from the weak damping assumption, to be in good agreement with their
experimental measurements at water-air and glycerol-air interfaces.
\changes{However, the results of Case A ($\beta=0.0625$) suggest that for
two-phase systems with bulk phases of similar fluid properties, such as
immiscible liquid-liquid flows, the frequency error ensuing from the weak
damping assumption is large enough to have a practical impact and to be
measured experimentally.}

The results presented in Fig.~\ref{fig:gammaNorm} suggest that, assuming
the same cumulative properties $\tilde{\rho}$, $\tilde{\mu}$ and $\tilde{\nu}$ of the two-phase
system, systems with high ratios of kinematic viscosity or density ($\beta
\rightarrow 0$) impose a smaller effective viscous damping than two-phase
systems in which the ratios of kinematic viscosity and density are unity ($\beta
= 0.0625$). Hence, two-phase systems with $\beta \rightarrow 0$, {\em e.g.}~Case
D, require a higher wavenumber to oscillate with a certain dimensionless
frequency $\hat{\omega}$ or to exhibit critical damping. This explains the observed shift to higher
wavenumbers in Eq.~(\ref{eq:criticalWavenumber}) for critical damping for
two-phase systems with smaller $\beta$.

\section{Conclusions}
\label{sec:conclusions}
In order to characterize the frequency dispersion of capillary waves in viscous
fluids, a rational parametrization based on a harmonic
oscillator model has been proposed, from which a formulation for the critical
wavenumber has been derived.
This critical wavenumber has been shown to be a characteristic value of the frequency dispersion of
capillary waves, as demonstrated by the consistent scaling of the undamped frequency
$\omega_0$ as well as the damping rate computed with AIVS and DNS for
representative two-phase systems. Critical damping occurs when capillary
and viscous timescales are in balance, a finding which may also apply to other
damped oscillators, such as elastic membranes, in that critical damping occurs when
suitably defined dispersive (or restoring) and dissipative timescales are in
balance.
 
The proposed scaling of capillary waves together with the critical wavenumber
obtained from the proposed harmonic oscillator model has been shown to lead to a
self-similar characterization of the frequency dispersion of capillary waves,
irrespective of the fluid properties and throughout the entire underdamped
regime.
The identified similarity of the frequency dispersion allows to take, for
instance, experimental measurements or results obtained with the AIVS of
\citet{Prosperetti1981} for certain fluid properties and translating these
results to any other two-phase system using the proposed scaling. Since
analytical solutions for simple reference cases are readily available, this
similarity, thus, enables an accurate {\em a priori} evaluation of the frequency
of a capillary wave in viscous fluids for any wavenumber.
Also, this similarity yields the conclusion that the wavenumber is a function of
the viscocapillary lengthscale and the property ratio, $k \sim
l_\mathrm{vc}^{-1} (1.0625-\beta)$, and the frequency is a function of the
viscocapillary timescale, $\omega \sim t_\mathrm{vc}^{-1}$.
\changes{Being able to accurately predict the dispersion of capillary waves for
arbitrary wavenumbers, can for instance help to better understand the
characteristics ({\em e.g.}~wavelength) of parasitic capillary waves riding on gravity waves or
the influence of viscous stresses on the capillary-driven breakup of liquid
jets, processes which are governed by nonlinear interactions between various
governing mechanisms.}

The inviscid and weak damping assumptions, which are commonly used to describe
the dispersion of capillary waves, have been shown to be inaccurate for high
wavenumbers, close but below the critical wavenumber. With respect to the
assumption of inviscid fluids, the presented results suggest this to be valid
for wavenumbers $\hat{k} \lesssim 10^{-3}$.
Interestingly, the weak damping assumption provides a considerably more accurate
description of two-phase systems in which the bulk phases have large density and
viscosity ratios than for systems with similar bulk phases. A new definition of
the damping rate based on the AIVS and DNS results has been proposed, which has
been shown to yield a more accurate prediction of the frequency of capillary
waves at high wavenumbers than estimates based on the inviscid and weak damping
assumptions.

\begin{acknowledgements}
The author acknowledges the financial support from the Engineering and Physical
Sciences Research Council (EPSRC) through Grant No.~EP/M021556/1. 
Data supporting this publication can be obtained from 
\href{http://dx.doi.org/10.5281/zenodo.58232}{http://dx.doi.org/10.5281/zenodo.58232}
under a Creative Common Attribution license.
\end{acknowledgements}

\end{document}